**Title:** Evaluating and improving real-world evidence with Targeted Learning

**Authors:** Susan Gruber[a], Rachael V. Phillips[b], Hana Lee[c], John Concato[d], Mark van der Laan[b]

[a]Putnam Data Sciences, LLC, Cambridge, MA
[b]University of California at Berkeley, Division of Biostatistics, Berkeley, CA
[c]Office of Biostatistics, Center for Drug Evaluation and Research, U.S. Food and Drug Administration, Silver Spring, MD
[d]Office of Medical Policy, Center for Drug Evaluation and Research, U.S. Food and Drug Administration, Silver Spring, MD

**ACKNOWLEDGMENTS / FUNDING STATEMENT**
This project was funded by the United States Food and Drug Administration (US FDA) pursuant to Contract 75F40119C10155. The content is the view of the author(s), and does not necessarily represent the official views of, nor an endorsement, by FDA/HHS, or the U.S. Government.

**CONFLICT OF INTEREST**
The authors report no conflicts of interest.

**ETHICS STATEMENT**
This manuscript is original work that reflects the views of all the authors. It should not be construed to represent the view of the FDA. No data were collected in connection with this manuscript.

**PATIENT CONSENT STATEMENT**
N/A

**PERMISSION TO REPRODUCE MATERIAL FROM OTHER SOURCES**
N/A

**ORCID**
Susan Gruber, https://orcid.org/0000-0003-0127-0099
Rachael V. Phillips, https://orcid.org/0000-0002-8474-591X
Hana Lee, https://orcid.org/0000-0001-9497-0574
John Concato, https://orcid.org/0000-0003-2846-0503
Mark van der Laan, https://orcid.org/0000-0003-1432-5511

# ABSTRACT

Purpose: The Targeted Learning roadmap provides a systematic guide for generating and evaluating real-world evidence (RWE). From a regulatory perspective, RWE arises from diverse sources such as randomized controlled trials that make use of real-world data, observational studies, and other study designs. This paper illustrates a principled approach to assessing the validity and interpretability of RWE.

Methods: We applied the roadmap to a published observational study of the dose-response association between ritodrine hydrochloride and pulmonary edema among women pregnant with twins in Japan. The goal was to identify barriers to causal effect estimation beyond unmeasured confounding reported by the study's authors, and to explore potential options for overcoming the barriers that robustify results.

Results: Following the roadmap raised issues that led us to formulate alternative causal questions that produced more reliable, interpretable RWE. The process revealed a lack of information in the available data to identify a causal dose-response curve. However, under explicit assumptions the effect of treatment with any amount of ritodrine versus none, albeit a less ambitious parameter, can be estimated from data.

Conclusion: Before RWE can be used in support of clinical and regulatory decision-making, its quality and reliability must be systematically evaluated. The TL roadmap prescribes how to carry out a thorough, transparent, and realistic assessment of RWE. We recommend this approach be a routine part of any decision-making process.

# Key Points

- Real world evidence (RWE) arises from diverse sources, including randomized controlled trials (RCT) that make use of real-world data, observational studies, and other designs.
- The Targeted Learning roadmap provides a systematic process for establishing whether the RWE provides transparent, reliable, actionable support for decision making.
- Stepping through the roadmap helps identify when data are not fit for purpose and pinpoints deficiencies in the data or study design.
- Outside of a regulatory environment, identifying barriers to causal effect estimation can suggest amelioration strategies that allow a different causal parameter to be learned from the available data.
- A thorough, realistic assessment of RWE using the Targeted Learning roadmap can become routine practice.

# 1. Introduction

From a regulatory perspective, real-world evidence (RWE) arises from diverse sources, including randomized controlled trials (RCT) that involves analysis of real-world data (RWD), observational studies, and other study designs [1]. RWE can provide insight into treatments, outcomes, and populations beyond those that can be studied in traditional RCTs. Researchers studying causal inference have established a strong theoretical foundation for understanding when and how causal effects can be estimated from RWD and developed sophisticated tools for doing so [2-4]. Strategies for increasing acceptance of RWE by improving its quality and promoting transparency have appeared in the literature [5-8]. The adage "trust, but verify" reminds us that before RWE can be used in support of clinical and regulatory decision-making, its quality and reliability must be systematically evaluated, from study design and conduct through analysis and interpretation.

Originally introduced as a guide for statistical learning from data, the Targeted Learning (TL) roadmap is also invaluable for establishing the validity and interpretability of findings from a RWD study (Figure 1) [9-11]. This paper demonstrates how to methodically step through the roadmap to expose weaknesses in causal claims. The roadmap provides a systematic way to evaluate the quality of RWE for both regulators and industry scientists. It can also inspire remediation strategies that strengthen the quality of the RWE.

Step 0. Well-defined scientific question, and precise description of the experiment generating the observed data
Step 1. Define a realistic statistical model, $\mathcal{M}$, in terms of the probability distribution of the data
Step 2. Define a causal model, and specify a causal quantity of interest, $\psi^{causal}$
Step 3. Determine statistical estimand in observed data, $\psi^{obs}$, that best approximates the causal quantity
Step 4. Estimation from data, respecting $\mathcal{M}$, and inference
Step 5. Interpretation and substantive conclusion supported by sensitivity analyses

Figure 1. The targeted learning roadmap.

A published retrospective cohort study will serve to illustrate how to detect and overcome insufficiency for causal effect estimation. Our intent is not to provide a commentary on the published findings, but to discuss concepts in causality and present results from an alternative data analysis. *Shinohara, et. al.* studied the association between ritodrine hydrochloride and maternal pulmonary edema in twin pregnancy in Japan [12]. Ritodrine had previously been shown to increase risk of pulmonary edema in pregnant women [13]. Study authors wanted to establish this result in the sub-population of women pregnant with twins, who are at higher risk of pre-term labor. In Japan, ritodrine is a first line therapy for halting pre-term labor, although in the United States, it was withdrawn from the market in 1995 due to efficacy and safety concerns [12, 14].

The target of the primary analysis was the dose-response association. The odds ratio (OR) for developing pulmonary edema associated with a one unit increase in total ritodrine dosage was estimated as OR = 1.02 (95% CI = (1.004, 1.03)). The authors state that due to unmeasured confounding, a causal interpretation is not warranted. The finding was interpreted as a partially adjusted measure of the dose-response association, since certain pre-existing health conditions that confound the treatment-outcome association were not available to the study team.

The study provides a rich example of challenges to learning from data, well beyond unmeasured confounding. In the next section we follow the TL roadmap to identify additional barriers to evaluating a causal dose-response curve. Subsequently, we discuss potential solutions that are based on specifying a statistical model that respects the process that gave rise to the data, crafting a realistic definition of treatment consistent with real-world feasibility, and selecting an alternative, less ambitious, target parameter. Results of a modified data analysis support the conclusion that ritodrine treatment increases risk for pulmonary edema.

## 2. Methods
### 2.1 Evaluating real-world evidence

Data made publicly available on Dryad by the study authors consists of observations on $n = 225$ women in Japan pregnant between 2009 and 2016 [15]. Each observation contains baseline covariates, $L(0)$; ritodrine treatment administered over multiple time points, $A(t)$; time-varying covariates at multiple time points, $L(t)$; the pulmonary edema outcome, $Y$; and additional covariates measured up to 24 hours post-delivery, $L(t+)$. Actual infusion rates of the study drug varied from 50 milligrams per minute (mg/min) to 200 mg/min, over a variable period of time. Total dosage was converted to units of 72 mg/24 hours. We step through the roadmap to better understand circumstances under which causal effect estimation might be possible, given the information available.

#### 2.1.1 Step 1. Statistical Model
Study authors posed a main terms logistic regression model, $logit\{P(pulmonaryEdema)\} = \beta_0 + \beta_1 ritodrine + \beta_2 PIH + \beta_3 BMI + \beta_4 PPH + \beta_5 corticosteroids + \beta_6 Mg + \beta_7 transfusion + \beta_8 term + \beta_9 bedrest$, where covariates are defined as
*ritodrine (the treatment)*: total dosage of ritodrine; *PIH*: pregnancy-induced hypertension (Y/N); *BMI*: body mass index (kg/m$^2$); *PPH*: postpartum hemorrhage; *term*: term birth (Y/N); *bedrest*: bed rest > 6 weeks (Y/N).

This defines the statistical model narrowly, in a way that almost certainly precludes the true data distribution. Including treatment in the model as a continuous main term automatically imposes a linear and monotonic dose-response relationship. Making this restrictive modeling assumption at the outset for all situations is unwarranted . In fact, the paper provides the crude proportion of outcome events observed in the RWD, grouped by dosage levels [12, Table 2]. Plotting these values suggests the dose-response relationship is, in fact, non-monotonic (Figure 2). The crude risk increases as total dosage approaches 50 units, then decreases at larger doses. Although adjusting for measured confounders might explain away some of the crude dose-response association, a main terms logistic dose-response model appears to be unrealistic.

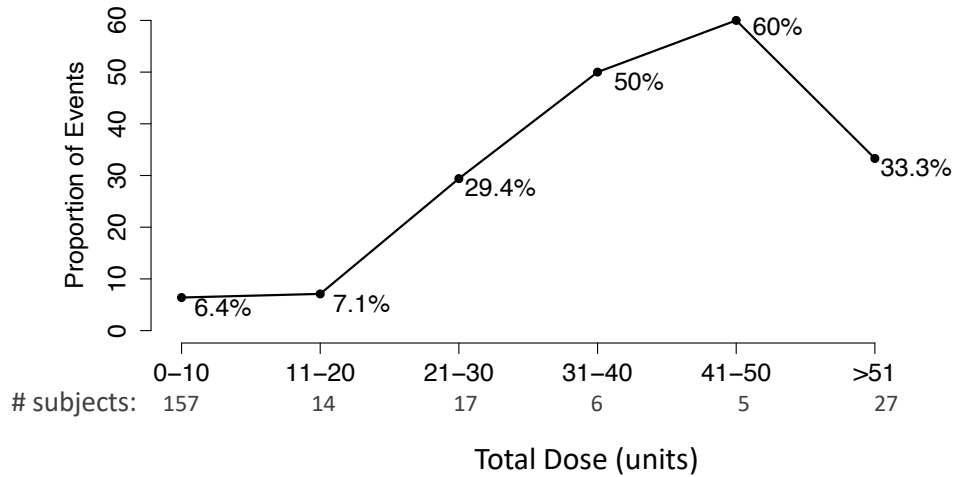

Figure 2. Proportion of patients with pulmonary edema grouped by total dose of ritodrine.

From a causal perspective, the timing of the outcome relative to other covariates included in the model is also problematic. The outcome event was measured from beginning of follow-up through 24 hours postpartum [S. Shinohara, personal communication, December 2019]. That means that at least two covariates, *PPH* (postpartum hemorrhage) and *term* (term birth), occurred after the outcome, for some women. Including post-outcome covariates in a causal dose-response model violates the tenet that a cause must precede an effect.

### 2.1.2. Steps 2 and 3: Causal estimand and corresponding statistical parameter
Because the causal model must be contained within the statistical model, here it is identical to the statistical model. Both the causal estimand and the statistical parameter are given by $\beta_1$.

### 2.1.3. Steps 4 and 5: Estimation and inference
Maximum likelihood estimates of the model coefficients, standard error (SE) estimates, and 95% confidence intervals (CIs) were calculated using standard methodology.

### 2.1.4. Step 6: Interpretation
Mathematically, the model coefficients quantify the projection of the true dose-response curve onto the model. However, because the model is highly misspecified, as discussed in Step 1, the parameter estimate is not equivalent to the causal conditional log odds associated with a one unit increase in total dose.

## 2.2 Strategies for improving the quality of real-world evidence

Deficiencies in the study design, model specification, and available data undermine confidence in any causal interpretation of the study finding, and the RWE generated by the study is arguably not reliable. Designing and carrying out a new study may not be feasible, but instead we can revisit the roadmap to see if it may be possible to learn something relevant from the data we have.

### 2.2.1 Step 1: Statistical Model

Parametric model misspecification can be avoided by defining a realistic, less restrictive, statistical model, $\mathcal{M}$. We define $\mathcal{M}$ non-parametrically as all distributions of the data consistent with the process by which treatment, covariates, and the outcome arise over time. We can further restrict $\mathcal{M}$ to distributions that respect the study inclusion criteria. This specification of the statistical model includes some distributions of the data where the dose-response relationship is monotonic and some distributions where it is not.

### 2.2.2. Step 2: Causal Estimand

The causal model makes conditional independence assumptions consistent with the time ordering of the data and assumes exogenous errors. The causal dose-response relationship can be defined in terms of a marginal multi-dimensional parameter. For example, given clinically meaningful groupings of total dosage administered, the mean risk for each treatment group can be targeted. Consider seven treatment categories: patients who receive no treatment ($A = 0$), or treatment at one of six levels ($A = 1, ..., 6$, corresponding to >0-10, 11-20, 21-30, 31-40, 41-50, 51+ units) (Figure 2). The causal dose-response parameter is written as $\psi^{causal} = (\psi_0, \psi_1, \psi_2, \psi_3, \psi_4, \psi_5, \psi_6)$, where $\psi_a$ is the counterfactual mean outcome observed if, contrary to fact, each patient received treatment at dosage level $A = a$. Furthermore, any causal contrast, such as the risk difference (RD), risk ratio (RR), or OR, can be easily calculated, e.g., $\psi_1 - \psi_0$ is the RD for treatment with up to 10 units of ritodrine vs. no treatment.

### 2.2.3 Step 3. Statistical estimand and assessment of identifiability

Next, we specify a statistical estimand in observed data, $\psi^{obs}$, that corresponds to our multi-dimensional $\psi^{causal}$ under identifying assumptions. For each dimension, $\psi^{obs}_{A=a} = E(Y|A = a, \bar{L}(t))$, where $\bar{L}(t)$ is the complete covariate history from baseline ($t = 0$) through the time the event occurred, or 24 hours post-delivery.

A problem is that the relative timing of $L(t)$ and $A(t)$, is not clearly recorded in the dataset. In other words, it is impossible to properly define $\bar{L}(t)$, thus we cannot specify any statistical parameter that corresponds to $\psi^{causal}$. Another complication is that clinicians may have slowed or stopped ritodrine infusion upon observing pulmonary edema in the patient. Under this scenario, the outcome partly causes the total dose, rather than the total dose causing the outcome. For these reasons, a causal dose-response curve is simply not identifiable from the data. Any study finding would rest entirely on a foundation of unrealistic modeling assumptions, and this RWE would not be appropriate to support decision-making.

*Alternative formulation:* In the absence of additional data that are fit for purpose, an alternative, unplanned analysis might provide insight into the causal relationship between ritodrine and pulmonary edema. Consider a simpler question: *does treatment with any dose of ritodrine vs. no treatment increase risk for pulmonary edema?* From this point treatment standpoint, the data consists of $n$ i.i.d. observations $O = (Y, A, W)$, where $Y$ is a binary outcome indicator, $A$ is a binary treatment indicator ($A = 1$ for treated, $A = 0$ for no treatment), and $W$ is a vector of baseline covariates. We are interested in a less ambitious causal parameter, the risk difference (RD). Downstream covariates that affect treatment infusion over time are irrelevant, and time-dependent confounding is no longer an issue. In the next subsection we step through the roadmap with this revised clinical question in mind.

## 2.3. Determining a point treatment effect by following the Targeted Learning roadmap

### 2.3.1 Step 1: Statistical Model

The statistical model is defined as all probability distributions of the data, with structure $O = (Y, A, W)$, consistent with study inclusion criteria.

### 2.3.2 Step 2: Causal Estimand

The causal model makes no further assumptions, beyond exogenous errors. The causal parameter of interest is the marginal RD, defined in terms of counterfactual outcomes by $\psi^{causal} = E(Y_1 - Y_0)$, where $Y_1$ is the counterfactual outcome under any level of treatment and $Y_0$ is the counterfactual outcome under no treatment.

### 2.3.3 Step 3: Statistical estimand and assessment of identifiability

The statistical estimand, $\psi^{obs} = E[E(Y|A = 1, W) - E(Y|A = 0, W)]$, has a valid causal interpretation when underlying assumptions are met [16]. The *consistency* assumption states that for each observation, the outcome under the observed exposure is equivalent to the counterfactual outcome that would be seen under that treatment assignment. It is satisfied under our simpler definition of treatment as any exposure to ritodrine, versus none.

The *positivity* assumption states that within all strata of confounders patients have a positive probability of receiving treatment at all levels considered. An outcome-blind look at the data shows that in some age groups no individuals were treated with ritodrine (Table 1), thus the parameter is not identifiable from the data. However, if coarser age categories are clinically justified, the violation can be eliminated by re-defining the age categories.

There is also another, more serious, violation of the positivity assumption. The prescribing information for ritodrine precludes administration when patients have serious pre-existing conditions, including maternal cardiac disease, hyperthyroidism, diabetes, and others [17]. If physicians adhere to these prescribing instructions, then no patients with these conditions would receive ritodrine, and the causal contrast in this subpopulation of women cannot be evaluated. However, these covariates aren't in the publicly available dataset. If the information was also not known to clinicians, then women with these conditions could possibly receive treatment, and there would be no theoretical violation of the positivity assumption.

Table 1. Number of subjects in control and treated groups by age in the original study age groupings (left), and re-defined age groupings (right).

| Original Categories | | | Re-defined Categories | | |
| --- | --- | --- | --- | --- | --- |
| Age | Control | Treated | Age | Control | Treated |
| 16-20 | 2 | 0 | 16 - 30 | 48 | 33 |
| 21-25 | 9 | 7 | | | |
| 26-30 | 37 | 26 | | | |
| 31-35 | 50 | 29 | 31-35 | 50 | 29 |
| 36-40 | 38 | 19 | 36-50 | 45 | 20 |
| 41-45 | 4 | 1 | | | |
| 46-50 | 3 | 0 | | | |

The final causal assumption, *coarsening at random* (CAR), is an assumption of no unmeasured confounders. With respect to the pre-existing conditions, if clinicians were unaware of patients' status, then these covariates could not have affected treatment decisions, so none are confounders. If clinicians were aware, then all of these covariates are unmeasured confounders.

*Alternative formulation:* One option would be to augment the exclusion criteria to rule out pregnant women who are ineligible to receive the study drug. The causal parameter would address a modified scientific question: *What is the marginal effect of ritodrine compared with no ritodrine on risk of pulmonary edema among women pregnant with twins to whom ritodrine may be prescribed?"* The RD would be identifiable from data, and interpretable as a subgroup-specific causal effect. Unfortunately, this approach isn't feasible with the available dataset, because we cannot identify patients with the relevant pre-existing conditions.

A second option would be to modify the scientific question again. Suppose we were interested in understanding how incidence of pulmonary edema would be affected if ritodrine were withdrawn from the market. The target population includes all women pregnant with twins, even those who are ineligible to receive ritodrine. The following *realistic treatment rules* [18] can always be followed,
> Rule 1: Treat with ritodrine unless expressly contra-indicated,
> Rule 2: Never treat with ritodrine

The marginal RD of pulmonary edema for following Rule 1 vs. Rule 2 can be estimated from observed data.

### 2.3.4. Step 4. Estimation and inference

Targeted minimum loss-based estimation (TMLE) with super learning (SL) was used to estimate the RD for the two realistic treatment rules. Potential baseline confounders included in the adjustment set were *age, height, weight, BMI,* and binary indicators of the following variables:

*obesity (BMI ≥ 25), first pregnancy, single placenta, assistive reproductive technology use, magnesium administration, and corticosteroid administration*. Analyses were run using R (v4.0.2), and the *tmle* (v1.5.0-1) and *SuperLearner* (vx2.0-26) packages [19-21]. For SL, the number of cross validation folds, V, was set to 20 to account for the number of events [22]. The default library of algorithms for modeling the outcome included linear regression, Bayesian additive regression trees (BART, in *dbarts* v0.9-18)[23], and lasso (*glmnet* v4.0-2*)*[24]*.* The default library for modeling the propensity score (PS) included logistic regression, BART, and generalized additive models (*gam* v1.20) [25]. TMLE requires the PS (1-PS) to be bounded away from zero for treated and untreated subjects, respectively. We set this lower bound to 0.06, based on the formula $5/[\sqrt{n}\ln(n)]$, with $n$ = 225) [26]. Influence curve-based SEs were reported by the software.

## 3. Results

PS diagnostics allow us to understand the overlap between treatment and control groups. The C-statistic associated with the PS model (0.72), and the plot of the PS distributions within each treatment group (Figure 3) indicate reasonable overlap of treated and comparator groups. No PS values were extreme, so truncation had no impact. The estimated RD was $\hat{\psi}^{obs} = 0.21$ (SE, $\hat{\sigma}$ = 0.062; 95% CI = [0.09, 0.33]).

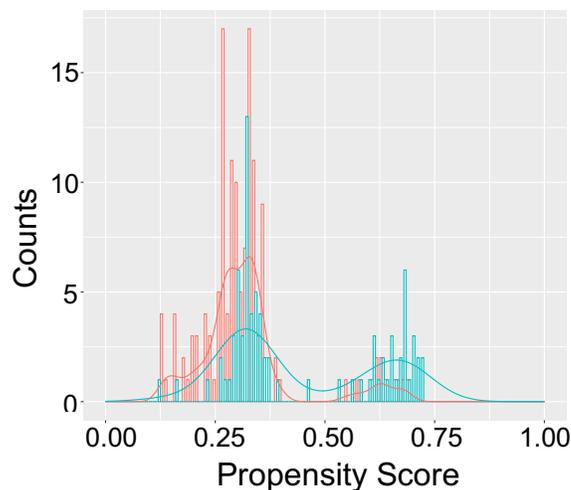

Figure 3. Distribution of propensity scores in treated (blue) and comparator (red) groups.

### 3.2 Step 5. Interpretation of the study finding and sensitivity analyses

Ritodrine was estimated to increase risk for pulmonary edema by 21%. Next, we assess if unexplained departures from the underlying causal assumptions could reverse that conclusion. If without our knowledge any of the three causal assumptions were violated, even at infinite sample size the estimated RD would not equal the true causal effect. The difference between the statistical parameter and the causal parameter is termed the *causal gap*, $\delta = \psi^{stat} - \psi^{causal}$. A non-parametric sensitivity analysis illustrates how point estimates and CI bounds change at different assumed values of $\delta$ [27].

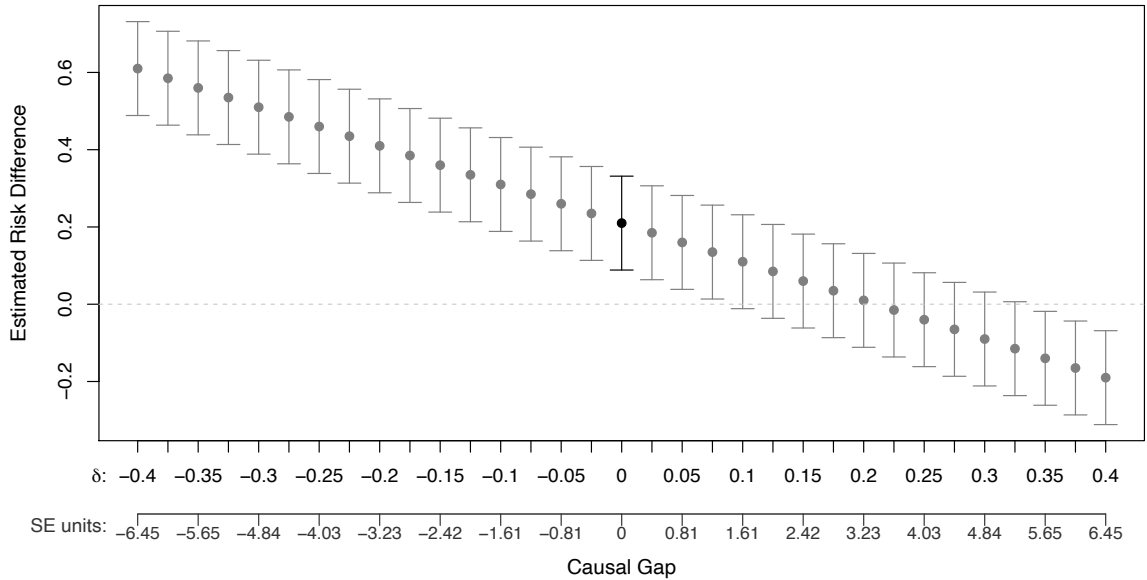

Figure 4. Non-parametric sensitivity analysis showing the risk difference and 95% confidence intervals under different presumed values of the causal gap, $\delta$, and also in standard error (SE) units.

Figure 4 shows the estimated RD and 95% CI under different values of $\delta$. For comparison, the size of the causal gap is also expressed in terms of the standard error of the effect estimate (SE units). The point estimate and 95% CI bounds determined from the data analysis are plotted at 0 on the x-axis. If there is a non-zero causal gap, then the estimate and CI would shift either left or right, depending on the direction and magnitude of the gap. Estimates and CIs plotted in gray correspond to different hypothetical causal gaps. If subject matter experts believe that any potential causal gap is likely to be negative then this sensitivity analysis reinforces a conclusion that ritodrine increases risk for pulmonary edema. If the causal gap is thought to be in the positive direction, unless the gap size is greater than approximately 0.1 the conclusion remains unchanged. The causal gap would have to be extremely large ( > 0.325) to conclude that ritodrine is protective for pulmonary edema.

## 4. Conclusions

For regulatory purposes, a well-designed study and good quality data are of paramount importance. Following the TL roadmap allowed us to systematically evaluate the suitability of these data for estimating a causal dose-response curve. Outside of a regulatory environment, the roadmap pointed us to explore alternative formulations of the causal question to produce more reliable, interpretable RWE.

Steps 1-3 of the roadmap crystallized the statistical learning task by defining the statistical model ($\mathcal{M}$), causal parameter ($\psi^{causal}$), and statistical estimand ($\psi^{obs}$), that can answer the corresponding clinical question of interest. The original study's overly restrictive $\mathcal{M}$ was ill-suited for modeling the true causal dose-response relationship. Our alternative formulation of a multi-dimensional $\psi^{causal}$ addressed that problem. However, given the uncertainty around the time ordering of treatment, covariates, and outcome, we were unable to describe a corresponding statistical estimand that could be identified from the data. Even without looking at the actual data, we were able to identify structural barriers that preclude evaluating a causal dose-response curve. This situation motivated targeting a point treatment parameter, defined in terms of realistic treatment rules.

Step 4, estimation of the statistical parameter, should go beyond fitting the coefficients in a single parametric model. If $\mathcal{M}$ is sufficiently general, i.e., realistic, then flexible machine learning (SL) is required. TMLE tailors the procedure for unbiased, efficient estimation of the statistical parameter, and provides influence curve-based inference.

Step 5, interpretation of the study finding, should incorporate a non-parametric sensitivity analysis that avoids imposing unwarranted parametric constraints. If a small, hypothetical but clinically plausible causal gap is sufficient to nullify or reverse the substantive conclusion, then the study findings are not a dependable guide decision making. On the other hand, when findings are robust in the face of plausible values of the causal gap, confidence is reinforced.

RWE can fulfill needs for information beyond that generated by RCTs. However, trust must be earned, not assumed. The TL roadmap provides a systematic process for establishing whether the RWE provides transparent, reliable, and actionable support for decision-making. A thorough, honest, realistic assessment of RWE can be a routine part of any decision-making process. The TL roadmap prescribes how this can be accomplished.